\documentclass[aip,jcp,reprint]{revtex4-1}
\usepackage{graphicx}
\usepackage{amsmath,amssymb}
\usepackage{subfig} 
\usepackage{xcolor}
\begin{document}
 
\title{Adsorption of polyelectrolytes on charged microscopically patterned surfaces}
 
 \author{Amin Bakhshandeh}
 \email{amin.bakhshandeh@ufrgs.br}
 \affiliation{Programa de  P\'os-Gradua\c c\~  ao em F\'\i sica\c, Instituto de F\'\i sica e Matem\'{a}tica, Universidade Federal de Pelotas, Caixa Postal  354, CEP 96010-900 Pelotas, RS, Brazil}
  \affiliation{Departamento de F\'\i sico-Qu\'\i mica, Instituto de Qu\'\i mica, Universidade Federal do Rio Grande do Sul, 591501-970, Porto Alegre, RS,
 	Brazil}

 \author{Maximiliano Segala}
 \email{maximiliano.segala@ufrgs.br}
 \affiliation{Departamento de F\'\i sico-Qu\'\i mica, Instituto de Qu\'\i mica, Universidade Federal do Rio Grande do Sul, 591501-970, Porto Alegre, RS,
 	Brazil}

\date{\today}

\begin{abstract}
In the present study we have investigated ,using Monte Carlo simulations (MC), the adsorption of polyelectrolytes on the charged nanopatterned surfaces. Different surface's patterns were considered and we noticed that the amount of adsorption is directly dependent on the size of the domains. Also in the case of checkerboard configuration it was observed that the polyelectrolytes are aligned along the diagonal of square domains.  
 
\end{abstract}

\maketitle

\section{Introduction}
Polyelectrolytes are polymers which possess ionized groups in polar solvents. There are many systems which are considered as polyelectrolyte such as
proteins, sulfonated and polyacrylic acids\cite{Barrat,BUDD}. 
They have a wide range of application from water treatment to tissue engineering~\cite{BOLTO,raj2018}.
Normally the "Click" chemistry is used as a powerful and efficient method to synthesis these systems, in this method, by carbon-hetero bond formation, polymers can be synthesized
\cite{LASCHE}. 
As these components in polar solvents are charged, the electrostatic interaction in these systems plays an important role\cite{dob,carrillo}. 
 
For example one of the important characteristics of polyelectrolytes is their extended chain, the main reason for this phenomena is Coulomb repulsion  
between charged segments\cite{levin,lowe}, the extended chain of polyelectrolytes, in turn, leads to a bigger hydrodynamic volume\cite{lowe}.
As we mentioned earlier, these components have a very wide range of application, therefore it is of critical importance to have a deeper understanding of properties and behavior of these components. 
One way to evaluate the physical and chemical properties of materials is to study their adsorption to different surfaces. By doing so, it can shed light on some material's behavior within a chemical process. In biological systems, molecules should adsorb to the surface of the Enzyme catalysis on  active sites, for this reason, the rate of reaction may depend on stability of the molecules on surfaces and also rate of adsorption and adherence.  As a result, we can understand the importance of interaction of molecules with surfaces. There are many surfaces which  gain charge inside polar solvents and since polyelectrolytes are charged  the electrostatic interaction becomes significance for adsorption of polyelectrolytes on these surfaces. 

At first glance, it may seem it is a simple problem because there are very primitive mean-field theories such as Deby-H{\"u}ckel 
approximation to explain the behavior of charged particles in aqueous solutions~\cite{Chodanowski,muthukumar,aubouy,levin,amin2018}. However, it is well 
known that theoretical study of adsorption of charged particles on charged plates is not easy since mean-field approximation collapses to describe their behavior in high correlated regions~\cite{amin2011,amin2018,levin}. Nevertheless, there are some approximations using simple models such as one component plasma near a neutralizing background or some similar 
theories which work well in regions of high electrostatic correlations. The situation gets worse when surfaces are not homogeneously charged or some nano-patterns exist on them. However, always one of the best ways to study these systems is computational simulations~\cite{von,netz,wallin,wallin2,McQuigg,wallin3,dos2016}. 
There are many biological systems which their surfaces carry patterns, as an example, proteins can 
bind to some significant patterns of surfaces~\cite{mcnamara,muthukumarpattern,muthukumarchain}.
Nowadays, it is easy to create charged nano-structured surfaces, these periodic organized surfaces are found in nanostructures, magnetic storage media and nanowires~\cite{velichko,seul,parthasarathy,piner}. 
The simulations of theses systems are more complicated than the simulation of simple uniform charged surfaces. McNamara et al.~\cite{mcnamara} studied adsorption of one polyelectrolyte on different patterned surfaces and approximated interaction between the charges on
the surface and polyelectrolyte's monomers with a screened Coulomb interaction which obtained from the Debye-H{\"u}ckel potential which is:
\begin{equation}
V(r) \propto \frac{e^{-k~r}}{r} ,
\label{eq1r}
\end{equation}
where $k$ and $r$ are inverse Debye length and interacting distance respectively.
 Eq.~\ref{eq1r} is a mean-field potential and as a result in their simulations the effect of neighbors cell was not considered. One way to produce different patterns is to put point charges on surface~\cite{amin2015}, the drawback of this method is that, depending  on the surface charge density and surface's size, one should put enough number of particles on the plate which this may lead to slow down the simulation.  Recently, an efficient and robust method has been introduced to simulate nanopatterned charged surfaces inside an electrolyte solution~\cite{amin}. In the mentioned method, to deal with the long-range Coulomb interaction between the ions a modified 3d Ewald summation method was used~\cite{yeh}. The surfaces are considered as periodic charged sinusoidal patterns. The analytical solution of the Poisson equation was evaluated in order to properly consider the effect of the nano-structured charged walls as an external potential. By using this method the adsorption of polyampholytes on these surfaces has been studied~\cite{amin2019}.

In the present work we will use Monte Carlo simulations to study the adsorption of Polyelectrolytes  on  charged nano-patterns surfaces
with the mentioned method. Our main aim is to observe the effect of charged patterns on adsorption of these macromolecules on nano-patterned surfaces.

The remainder of the paper is organized as follows.:
 In section~\ref{model}, we explain the model and the simulation details. In section~\ref{results}, we 
summarize our results. In section~\ref{conclusions} we conclude our work. 

\section{THE MODEL AND SIMULATION DETAILS}\label{model}

\begin{figure}[h]
\includegraphics[width=9cm]{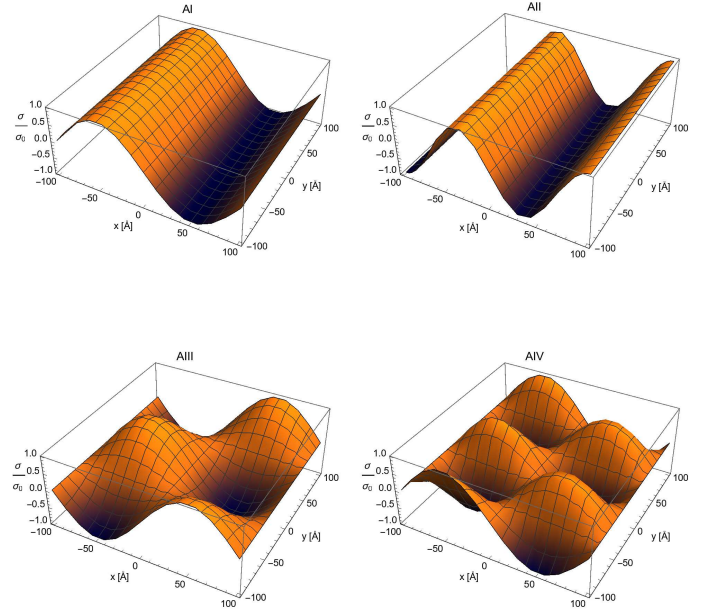}
\caption{Representation of $\sigma / \sigma_0$ against $x$ and $y$ direction for $\sigma_0 = 0.1~$C/m$^2$ and $\varphi_{x}=\varphi_{y}=0$. AI) $n_x =1$, $n_y =0$, AII) $n_x =2$, $n_y =0$, AIII) $n_x =1$, $n_y =1$, AIV) $n_x =2$, $n_y =2$.   }
\label{f1}
\end{figure}

We model nano-patterned surface by a sinusoidal charge density, which mathematically can be described as follows~\cite{amin}:
\begin{equation}
\sigma(x,y) = \sigma_0  \sin(k_x x+\varphi_x)  \sin(k_y y + \varphi_y) \ ,
\label{eq1}
\end{equation}
where $\sigma_0$ is the amplitude, $\varphi_{x}=0$ and  $\varphi_{y}=\pi/2$  are the phases, $k_x = 2 \pi n_x/L_x$  , $k_x = 2 \pi n_y /L_y$, with $L_x$ and $L_y$ periods
of charge density oscillations in $x$ and $y$ directions, respectively, and $n_{x,y} $ are integers. By changing $n_{x,y} $ different patterns simply can be generated.  We have plotted $\sigma / \sigma_0$ for four cases in Fig.\ref{f1}, where $\sigma_0 = 0.1~$C/m$^2$.

It can be shown that the potential produced by this charge density can be written as~\cite{amin}
\begin{equation}
\Phi_1 (\bold{r} )= \frac{2 \pi \sigma_0}{\epsilon_w \alpha}  \sin(k_x x + \varphi_x)  \sin(k_y y  + \varphi_y) e^{-\alpha \lvert z \rvert} \ ,
\label{eq3}
\end{equation}
where $\alpha = \sqrt{k_x^2 + k_y^2}$. We consider our system  consisting of two flat surfaces of dimensions $L_x$ and $L_y$, located at $z=-L/2$ and $z=L/2$ where $L=300$~$\AA$ enclosing the electrolyte solution. Also, we set $L_x=L_y=200~$~$\AA$~to have a better statistic . The solvent is assumed to be an uniform dielectric of permittivity $\epsilon_w$. The Bjerrum length is defined as $\lambda_B = e^2/k_BT\epsilon_w$ where $e$, $k_B$ and $T$ are the elementary charge, the Boltzmann constant and 
the absolute temperature, respectively. The Bjerrum length in current study is $7.2$~$\AA$, a value for room temperature and $\epsilon_w= 80$.

The electrostatic potential produced by both surfaces is given by~\cite{amin}
\begin{eqnarray}\label{ener}
\Phi(\bold{r} )= \frac{2 \pi \sigma_0}{\epsilon_w \alpha} \sin(k_x x + \varphi_x)  \sin(k_y y + \varphi_y)\nonumber \\ \left( e^{-\alpha \lvert z+L/2 \rvert} + e^{-\alpha \lvert z-L/2 \rvert} \right) \ .
\end{eqnarray}

The dissociated $100$ macromolecules between the surfaces are modeled with the primitive model. The monomers are modeled as hard spheres of radius $2~$\AA, with centered negative charge   $-e$. At first we consider  polyelectrolytes which are composed of $18$ monomers of charge $-e$. Since polyelectrolytes are overall charged, in order to make system neutral we add counterions to the systems.
The adjacent monomers that compose a chain interact via a parabolic potential as $U_b(r)=A/2(r-r_0)^2$, where $A=0.97~k_BT$, $r$ is the distance between adjacent monomers and $r_0=5~$\AA~\cite{dos2016}. The simulations are performed using the Metropolis algorithm~\cite{Frenkel,Allen}, with $5 \times 10^7$  MC steps for equilibration. Each sample is obtained with $300$ trial movements per particle. The macromolecules can perform rotation move and head and tail monomer can be exchange at each MC movement. Moreover, monomers can have short displacements, which models vibration of segments~\cite{dos2016}.  Since the system has slab geometry we use a corrected 3D Ewald summation~\cite{yeh}.
The total potential energy of the system composed of $N$ hard sphere particles of charge $q_i$ located at ${\pmb r}_i$ can be written as follows~\cite{dos2016,amin}:
\begin{eqnarray}\label{ener1}
U=\sum_{{\pmb k}\neq{\pmb 0}}^{\infty}\frac{2\pi}{\epsilon_w V |{\pmb k}|^2}\exp{[-\frac{|{\pmb k}|^2}{4\kappa_e^2}]}[A({\pmb k})^2+B({\pmb k} )^2] + \nonumber \\
\frac{2\pi}{\epsilon_w V} M_z^2 + \dfrac{1}{2}\sum_{i \ne j}^Nq_iq_j\frac{\text{erfc}(\kappa_e|{\pmb r}_i-{\pmb r}_j|)}{\epsilon_w |{\pmb r}_i-
{\pmb r}_j|}+\nonumber \\
\sum_{i=1}^{N}q_i\Phi ({\pmb r}_i )+\sum{\vphantom{\sum}}' U_b(|{\pmb r}_i-{\pmb r}_j|) \ ,
\end{eqnarray}
where
\begin{eqnarray}
A({\pmb k})&=&\sum_{i=1}^N q_i\text{cos}({\pmb k}\cdot{\pmb r}_i) \ , \nonumber \\
B({\pmb k})&=&-\sum_{i=1}^N q_i\text{sin}({\pmb k}\cdot{\pmb r}_i) \ , \nonumber \\
M_z&=&\sum_{i=1}^N q_i z_i \ , \nonumber \\
\end{eqnarray}
and $V=L_x \times L_y \times L_z$ is the volume of the main cell, while $L_z=3L_x$. The ${\pmb k}$ vectors are defined as ${\pmb k}=(\frac{2\pi}{L_x} n_x,\frac{2\pi}{L_y} n_y,\frac{2\pi}{L_z} n_z)$, where $n's$ are integers. Around $500$ vectors are used in the calculation. The damping parameter is $\kappa_e=5/L_x$. The restricted summation in the last sum in Eq.~\ref{ener1} is due to adjacent monomers in macromolecules.

\section{Results}\label{results}
 As we discussed, for many chemical and physical process it is great of importance that molecules adsorb on the surfaces, as a result, it is interesting to study the effect of patterns on adsorption.

\begin{figure}[t]
  	\includegraphics[width=9cm]{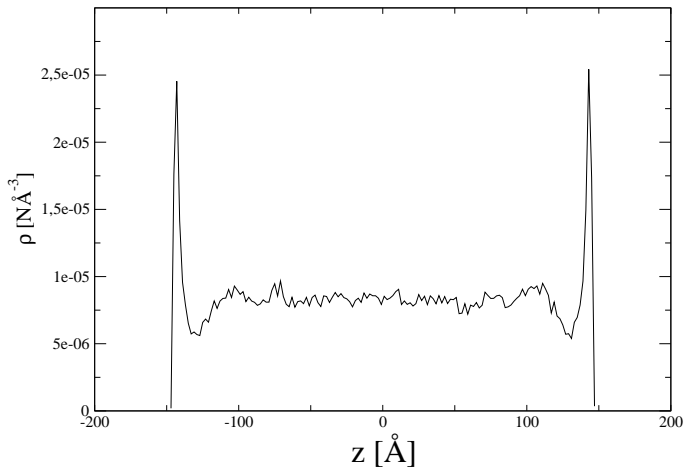}
  	\caption{Concentration profiles of polyelectrolytes (center of mass) for case of AI for polyelectrolytes with $18$ monomers.~$\sigma_0 =0.1$~C/m$^2$, $n_x=1$ and $n_y=0$ }
  	\label{fig12}
  \end{figure}

 We put $100$ polyelectrolytes in the cell in the presence of surfaces with different patterns. We have shown the density profile for $(n_{x},n_{y}) =(1,0)$  in Fig.~\ref{fig12}. 

\begin{figure}[t]
	\includegraphics[width=9cm]{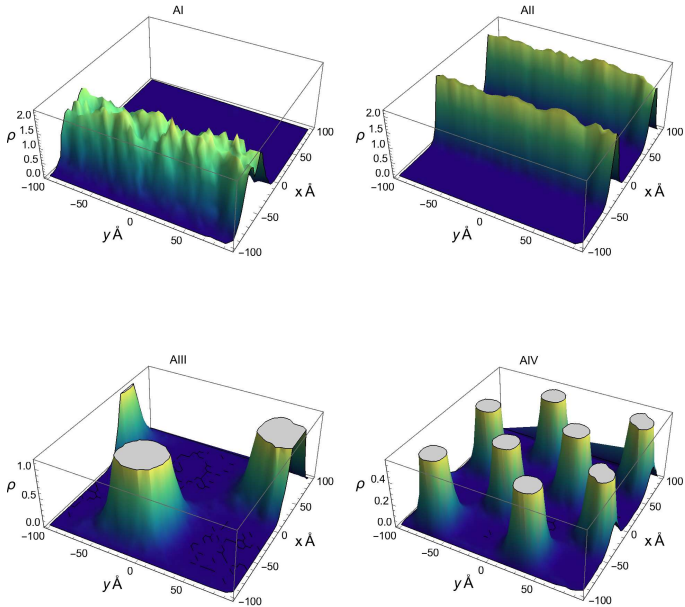}
	\caption{Density profile of segments of polyelectrolytes with $18$ monomer in a bin $\Delta z = 6 $ \AA at contact for
		AI)~$n_x = 1$ and $n_y = 0$, AII)~$n_x = 2$ and $n_y = 0$,AIII)~$n_x = 1$ and $n_y = 1$,AIV)~$n_x = 2$ and $n_y = 2$ respectively. }
	\label{f2}
\end{figure}

  As is seen in Fig.~\ref{fig12} there is high adsorption on surfaces. Since the charge of monomers is negative we expect to have more adsorption on domains with opposite charge of polyelectrolyte. 
  We have plotted the density profile of monomers near the plate's surface for different $(n_{x},n_{y})$, which are $(1,0)$, $(2,0)$, $(1,1)$ and $(3,3)$ respectively. As is seen in Fig~\ref{f2}, all polymers are adsorbed in domains with opposite charges, for the case of  $(1,0)$, as the positive region is bigger the density peak is wider, however for  $n_{x,y}$, $(1,1)$ the peaks become more sharper. The same phenomena is observed for $(3,3)$. 
  
\begin{figure}[t]
 \includegraphics[width=9cm]{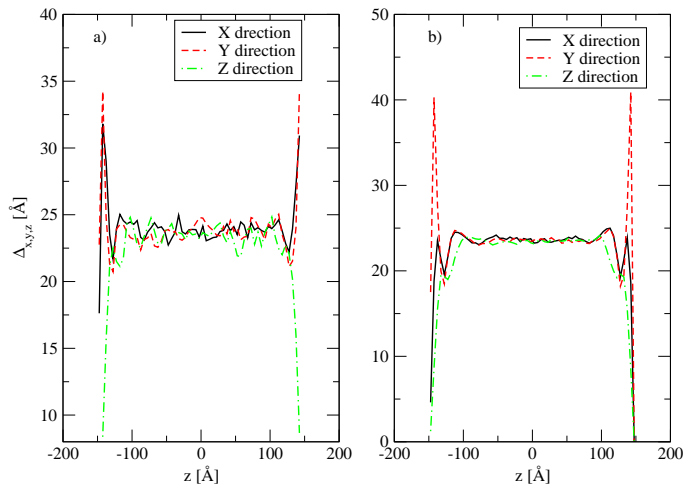}
 \caption{Average distance of between head to tail of polyelectrolytes for component in $x$, $y$ and $z$ direction for a) case AI and b)AII.}
 \label{fig4}
 \end{figure}

\begin{figure}[t]
	\includegraphics[width=9cm]{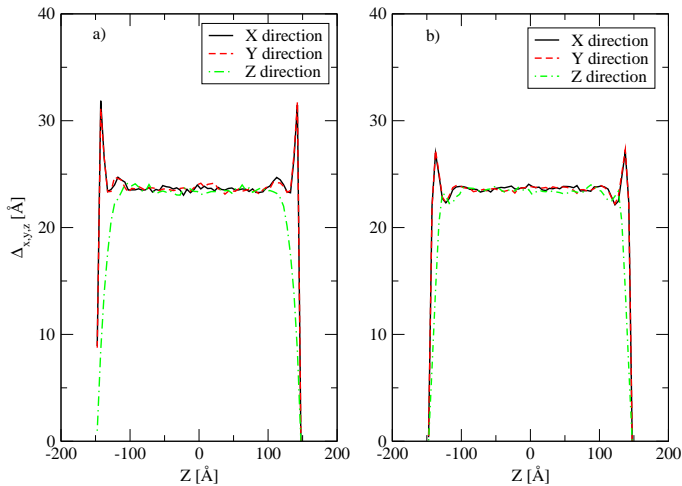}
	\caption{Average distance of between head to tail of polyelectrolytes for component in $x$, $y$ and $z$ direction for a )case AIII and b) AIV. }
	\label{fig5} 
\end{figure}

At this point, it is interesting to realize the geometrical configuration of adsorbed polyelectrolytes on different patterns.
To this end, we use the separation distance between the first and the last monomer of each polyelectrolyte, by doing this we can obtain a picture of shape of polymer in different positions in the cell, especially near the plate. 
The mentioned distances are defined as follows:
  \begin{eqnarray}\label{ener}
  \Delta x(z)= \sqrt{ \Bigg \langle \frac{\sum_{i=1}^{N}{ (x_{i}^{head}-x_{i}^{tail})^2}}{N(z)}\Bigg \rangle}   , \nonumber \\ 
  \Delta y(z)= \sqrt{ \Bigg \langle \frac{\sum_{i=1}^{N}{ (y_{i}^{head}-y_{i}^{tail})^2}}{N(z)}\Bigg \rangle}    , \nonumber \\ 
  \Delta z(z)= \sqrt{ \Bigg \langle \frac{\sum_{i=1}^{N}{ (z_{i}^{head}-z_{i}^{tail})^2}}{N(z)}\Bigg \rangle}  \ ,
  \end{eqnarray}
  where $\Delta x$, $\Delta y$  and $\Delta z$ are the rms components of the head-to-tail vector  in $x$,$y$ and $z$ directions and $N(z)$  
  is the number of polyelectrolytes in each volume element in position $z$.
  We have plotted head to tail distances for cases AI, AII, AIII and AIV in Figs.~\ref{fig4} and \ref{fig5}. 

 As is seen in Fig.~\ref{fig4}a for the case of $n_{x,y} = (1,0)$ the $\Delta z$ is more and less constant and both $\Delta x$ and $ \Delta y$ components are high near the surface in comparison with bulk value. Nevertheless,  $\Delta y$ component is a little bit higher than $\Delta x$ component, the reason is that polyelectrolytes prefer to extend along the positive region to have less repulsion from negative domains and have more attraction by positive domain. 
 
\begin{figure}[t]
	\includegraphics[width=9cm]{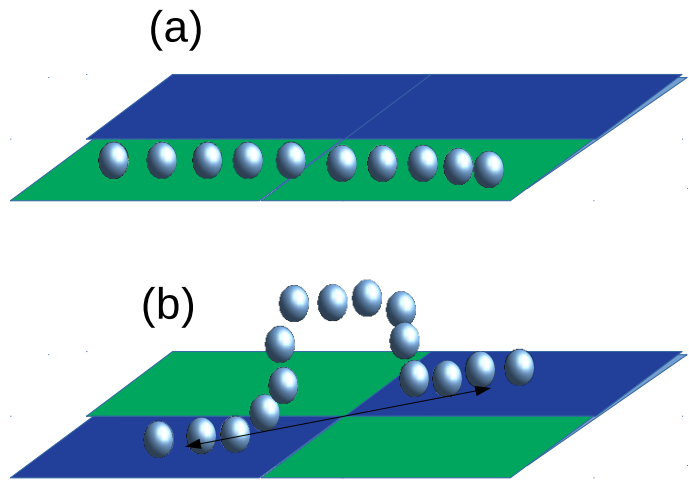}
	\caption{The schematic representation of polyelectrolytes adsorbed on the charged patterned surfaces. a) Stripes configurations b) Checkerboard configurations. The long polyelectrolytes try to reach other opposite charged domains via diagonal of domains by getting distance from plate near the same charged domains. }
	\label{fr3}
\end{figure}

 For the case of   $n_{x,y} = (2,0)$ as is seen in Fig.~\ref{fig4}b the $\Delta x$ component decreases dramatically and $ \Delta y$ component increases. This also can be interpreted by using the Fig.~\ref{f2}, since the positive area on the plate decreases and the negative monomers do not like to be in contact with negative regions in this case polyelectrolytes adjust themselves in $y$ direction, in order to be far away from negative regions on the plate. We have shown this configuration schematically in Fig.~\ref{fr3}a

\begin{figure}[t]
	\includegraphics[width=7cm]{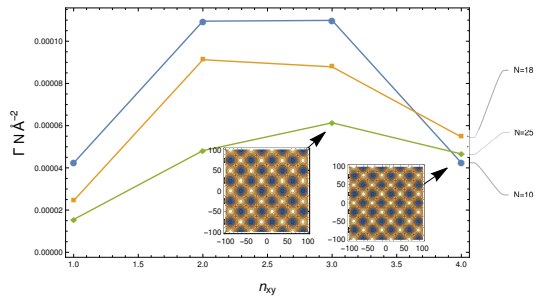}
	\caption{The adsorption against $n_x$ for polyelectrolytes with different number of segments $N_{segments}=10$, $18$ and $25$. The surface charge is normalized by $\sigma_0/n_x =0.1$ C/m$^2$. In all configurations $n_x = n_y$. }
	\label{figl} 
\end{figure}

\begin{figure}[t]
	\includegraphics[width=8cm]{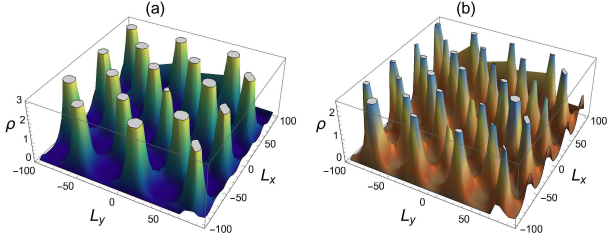}
	\caption{Density profile of segments of polyelectrolytes in a bin $\Delta z = 6 $ \AA at contact for polyelectrolyte with $25$ segments.
		a)~$n_x = 3$, $n_y = 3$, b)~$n_x = 4$ and $n_y = 4$.  }
	\label{fr}
\end{figure}

In next, we study the cases $n_{x,y} =(1,1)$ and $n_{x,y} =(2,2)$, as is observed in Fig.~\ref{fig5} for the case of $n_{x,y} =(1,1)$ and $n_{x,y} =(2,2)$, the values of $\Delta x$ and $\Delta y$ become equal near the plates, this is due to the fact that polyelectrolytes are aligned along the diagonal of domains, however for the case of AIV their values decreases, which this is because of smaller size of domain. 
  In order to realize how the size of the domains and polyelectrolytes affect on the adsorption, we obtain $\Delta R= \sqrt{\Delta x ^2 + \Delta y ^2+ \Delta z ^2}$ in bulk, by using MC simulation, for three different polyelectrolytes with number of segments $10$, $18$ and $25$ respectively. $\Delta R$  for polyelectrolytes with $10$, $18$ and $25$ is obtained $33$, $63$ and $80$ \AA. In next step, we implement MC simulation for different charge densities and patterns in the way that surface charge density gets scaled by $\sigma_0/n_x  = 0.1 $ C/m$^2$, also we put $n_x=n_y$ to create checkerboard configurations. The adsorption $\Gamma$ can be calculated by 
 \begin{equation}
 \Gamma=\frac{1}{2} \int_{-L/2}^{L/2}{(\rho(z) -\rho_b) dz}  ,
 \label{eq3}
 \end{equation}
 where $\rho(z)$ and $\rho_b$ are local and bulk density respectively.~$\rho_b$ is obtained by taking average of polyelectrolytes density at the middle of cell between $z=-25$ and $z=25$~\AA.
As is seen in Fig.~\ref{figl} when $\Delta R$ becomes comparable with domain's sides the adsorption starts to decrease, when the number of segments $N_{segments}$ is $10$  this happens at $n_x=n_y=4$ where domain's side, $l$,   is $25$~\AA~and when $N_{segments}$ is $18$ this happens at $n_x=n_y=3$ where $l =33$~\AA. We expect the same observation is seen for $N_{segments} = 25$ and decreasement in $\Gamma$ happens at $n_{x,y}=(2,2)$, however we observe that not only decreasement in $n_{x,y}=(2,2)$ but also  there is increment in $n_{x,y}=(3,3)$. In Fig.~\ref{fr} we have shown the density profiles of segments for two mentioned surface's configurations. As is seen for the case $(3,3)$ there is a wide peak at the center of each positive domain and for the case of $(4,4)$ the peaks become sharper which is due to repulsion force of other negative domains, however, between the peaks in diagonal directions, it is seen that there is a small density of segments which connect the peaks. This confirms the alignment of polymers in that direction, also the small magnitude of density in that regions suggests that the monomers stay in further distance from plates in comparison with other monomers. We show this configuration in Fig.~\ref{fr3}b. 

\begin{figure}[t]
	\includegraphics[width=8cm]{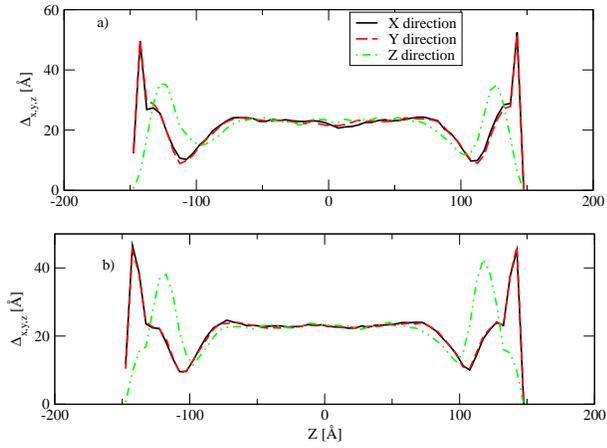}
	\caption{{Average distance of between head to tail of polyelectrolytes with $25$ segments for component in $x$, $y$ directions for a )$n_x=n_y=3$ and b)$n_x=n_y=3$.}   }
	\label{zfr1}
\end{figure}

\begin{figure}[t]
 	\includegraphics[width=8cm]{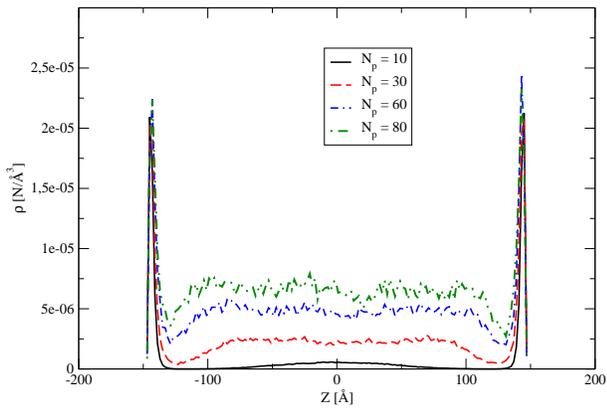}
 	\caption{Density profile of center of mass of polyelectrolytes with $10$ segments and different concentration inside the cell. $\sigma_0 =0.1$ C/m$^2$ and $n_x=1, n_y=0$} 
 	\label{fr1}
 \end{figure}

	 By looking at the Figs.~\ref{zfr1} it is seen that the $\Delta_x = \Delta_y$  which shows the same conclusion about the polymer configuration. This shows that despite the repulsion of other negative neighbor domains the polymer can  connect itself to other positive domains as is shown in Fig~\ref{fr3}b. However, this adjustment of polymer costs energy because in the region between domains the monomers feel more repulsion and want to avoid that region. As a result, the segments get distance from surfaces. By decreasing the edge of the domains the repulsion force becomes too strong that this special shape does not help polymer to gain more negative energy, as a result the adsorption decreases as we expected. In the end, we studied the effect of the concentration of polyelectrolytes on adsorption. To this end, we consider a different number of polyelectrolytes with $10$ segments inside the cell. We fixed the $\sigma_0 =0.1$ C
	 /m$^2$ and $n_{x,y}=(1,0)$. As can be seen in Fig.~\ref{fr1} by increasing the polymer concentration the adsorption increases however very rapidly the adsorption stop increasing due to saturation of the surfaces. 

\section{Conclusion}\label{conclusions}

In the present work, we have studied the adsorption of poly-
electrolytes to nano-patterned charged surfaces using a recent
method for simulation of these kinds of systems~\cite{amin}. It is shown that all polyelectrolytes are adsorbed to domains with opposite charge and concentrated at the center of the domains where there is higher charge density.
By studying the average distance of head to tail of molecules we observed that the molecules prefer to become extended on the surface in direction of $y$ for stripes configurations where the plate has transnational symmetry in that direction.
Also, it was observed that the amount of adsorption may relate to the head to tail distance of molecule in bulk. For checkerboard configurations, when the edge of the domain becomes comparable with this significance distance (head to tail distance) the adsorption decreases for the same scaling variable which is $\sigma_0/n_x$. However for longer polyelectrolytes since they can have a bend configurations along the diagonal of square domains and still have some segments near the opposite charged domains, they can find a more stable configuration and reach to the center of others opposite neighbor charged domains via diagonal path and as a result it can be seen that adsorption increases by reducing the size of domains. However for $n_x=n_y=4$ this configuration does not help to gain more attraction and stable configuration, as a result the adsorption reduces.  
\section{Acknowledgments}
 This work was supported by CAPES under process number 88882.306664/2013-01.
\clearpage
\bibliography{paper.bib}
\end{document}